# Single-molecule phosphorescence and intersystem crossing in a coupled exciton-plasmon system

Abhishek Grewal[1], Hiroshi Imada[2], Kuniyuki Miwa[3], Miyabi Imai-Imada[2], Kensuke Kimura[2], Rafael Jaculbia[2], Klaus Kuhnke[1], Klaus Kern[1,4], and Yousoo Kim[2]

[1]*Max-Planck-Institut für Festkörperforschung, Heisenbergstrasse 1, Stuttgart 70569, Germany*

[2]*Surface and Interface Science Laboratory, RIKEN, Wako, Saitama 351-0198, Japan*

[3]*Institute for Molecular Science, Myodaiji, Okazaki, Aichi 444-8585, Japan*

[4]*Institut de Physique, École Polytechnique Fédérale de Lausanne, Lausanne 1015, Switzerland*

---

ABSTRACT: Scanning the sharp metal tip of a scanning tunneling microscope (STM) over a molecule allows tuning the coupling between the tip plasmon and a molecular fluorescence emitter. This allows access to local variations of fluorescence field enhancement and wavelength shifts, which are central parameters for characterizing the plasmon-exciton coupling. Performing the same for phosphorescence with molecular scale resolution remains a significant challenge. In this study, we present the investigation of phosphorescence from isolated Pt-Phthalocyanine molecules by analyzing tip-enhanced emission spectra in both current-induced and laser-induced phosphorescence. The latter directly monitors singlet-to-triplet state intersystem crossing of a molecule below the tip. The study contributes to a detailed understanding of triplet excitation pathways and their potential control at sub-molecular length scales. Additionally, the coupling of organic phosphors to plasmonic structures is a promising route for improving light-emitting diodes.

## INTRODUCTION

Elucidating the processes occurring within isolated luminescent molecules has witnessed significant progress in recent years. Detailed studies have been carried out using STM metal-tip-enhanced molecular electrofluorescence[1-10] and photofluorescence[11-17] with atomic scale precision. Extending this experimental approach to phosphorescence with plasmonic tip enhancement has turned out to be a significant challenge because of low intensity and debated assignments of observed emission lines. Phosphorescence, the emission resulting from an excited triplet state transitioning to a ground singlet state, is a nominally forbidden transition and thus cannot be directly excited by light. However, this transition can become observable via "intensity borrowing" when triplet and singlet states mix, for example, through spin-orbit coupling (SOC). Excitation of triplet states occurs either by intersystem-crossing (ISC) from a higher-lying singlet state (which can be directly excited by light absorption) or by charge injection. ISC is a process operating on the molecular states in the opposite direction as up-conversion electroluminescence (UCEL), which has garnered increasing interest in single-molecule STM studies recently.[18-21] Moreover, molecular excitation through electron injection with random spin as employed in organic light emitting diodes (OLEDs) can generate spin-triplet ($T_1$) and spin-singlet ($S_1$) states in a 3:1 ratio.[22] Harnessing the otherwise lost energy of the triplet state occupancy is a major motivation for incorporating organic phosphors into host layers of electroluminescent OLEDs, kick-starting the search for suitable triplet emitters.[23] Recently, decay rate enhancement via coupling to plasmonic systems[24-26] in OLEDs has shown improved device stability[27] against ageing.[28, 29] For this reason, studying the influence of plasmon coupling on relative fluorescence and phosphorescence intensities, as well as ISC, at the single-molecule level is of great interest.

In this context, it is surprising that the direct observation of phosphorescence in close proximity to a metallic tip remained elusive for a long time. Early studies addressing this issue remain under debate, as the emission line was attributed either to phosphorescence[30] or to fluorescence from a charged molecular state[8, 16, 31]. A subsequent investigation[32] revealed that both states lie extremely close in energy, preventing definitive assignment. Questions also persist regarding the relative intensities of "forbidden" phosphorescence with respect to dipole-allowed transitions occurring between energetically higher-lying states. Due to the longer lifetime of the triplet state, its branching ratio to radiative decay is assumed negligible for a molecule in an STM and thus in proximity to a conductor. However, there is consensus that the lowest triplet state plays a very crucial role as a relay state in UCEL[20, 21, 33] and is essential for a comprehensive understanding of intramolecular excitation dynamics.[34]

To avoid ambiguities, we present a study of a molecule known for its strong phosphorescence emission. Using an STM (operated at a temperature of 4.3 K in vacuum at pressures < $10^{-11}$ mbar) equipped with an optical detection system that enables tip-enhanced optical spectroscopy with atomic resolution. A low coverage of molecules of platinum (II) phthalocyanine (PtPc) is evaporated in ultra-high vacuum (UHV) on top of few monolayers (ML) of NaCl epitaxially grown on an Ag(111) surface. The NaCl layer reduces the luminescence quenching that may occur due to hybridization with substrate electronic states.[35] PtPc is a well-suited candidate for this study due to its intense phosphorescence[36-39], attributed to strong SOC[37] facilitated by its heavy central Pt atom. The molecule provides an excellent platform for studying exciton-plasmon (X-P) interaction[2, 9, 40] in conjunction with ISC and phosphorescence, leading to a favorable condition for the observation of phosphorescence.[37, 41-43] In this study, we report strong fluorescence from PtPc and also investigate its phosphorescence, which remained previously unobserved in low-temperature STM experiments.[33] We further investigate how the properties of fluorescence and the coupling to plasmons modify the observed phosphorescence intensity. Finally, we provide clear evidence of ISC at the single-molecule level using tip-enhanced photoluminescence[11, 12], while avoiding the direct excitation of phosphorescence through charge injection into the molecule.

## RESULTS AND DISCUSSION

### STM-induced electroluminescence (STML) of a single phosphor

For details of sample preparation and experimental procedures see the Supporting Information 1.1. PtPc molecules adsorb in planar geometry on NaCl, with their four isoindole units aligned along the [100] and [010] directions of the NaCl lattice (Fig. 1A), and with the metal center positioned at a Na site (inset of Fig. 1A). Differential conductance (d$I$/d$V$) spectra of the molecule reveal molecular resonances with onsets at -2.2 V and +0.8 V (marked by arrows in Fig. 1B), corresponding to positive and negative ion resonances of PtPc. These resonances represent tunneling channels through the highest occupied (HOMO) and the degenerate lowest unoccupied molecular orbitals (LUMO and LUMO+1), respectively. The apparent transport gap of ~3.1 eV observed for PtPc/4-ML NaCl/Ag(111) agrees with prior experimental studies[33, 44-46] and theoretical simulations (vertical gray lines in Fig. 1B) performed using the Pauli master equation methods combined with first-principle calculations based on the density functional theory (DFT) (see Supporting Information 1.3 and 1.4). Constant height d$I$/d$V$ maps (insets in Fig. 1B) image the molecular frontier orbitals.

Figure 1C illustrates the STML experimental setup (see Supporting Fig. 2 for the optical setup, including the extension for TEPL). Figure 1D presents a Perrin-Jabłoński diagram detailing molecular emission processes triggered by electronic excitation through charge injection into the frontier molecular orbitals. This results in the direct excitation of both $S_1$ and $T_1$ states, responsible for fluorescence and phosphorescence emission, respectively.

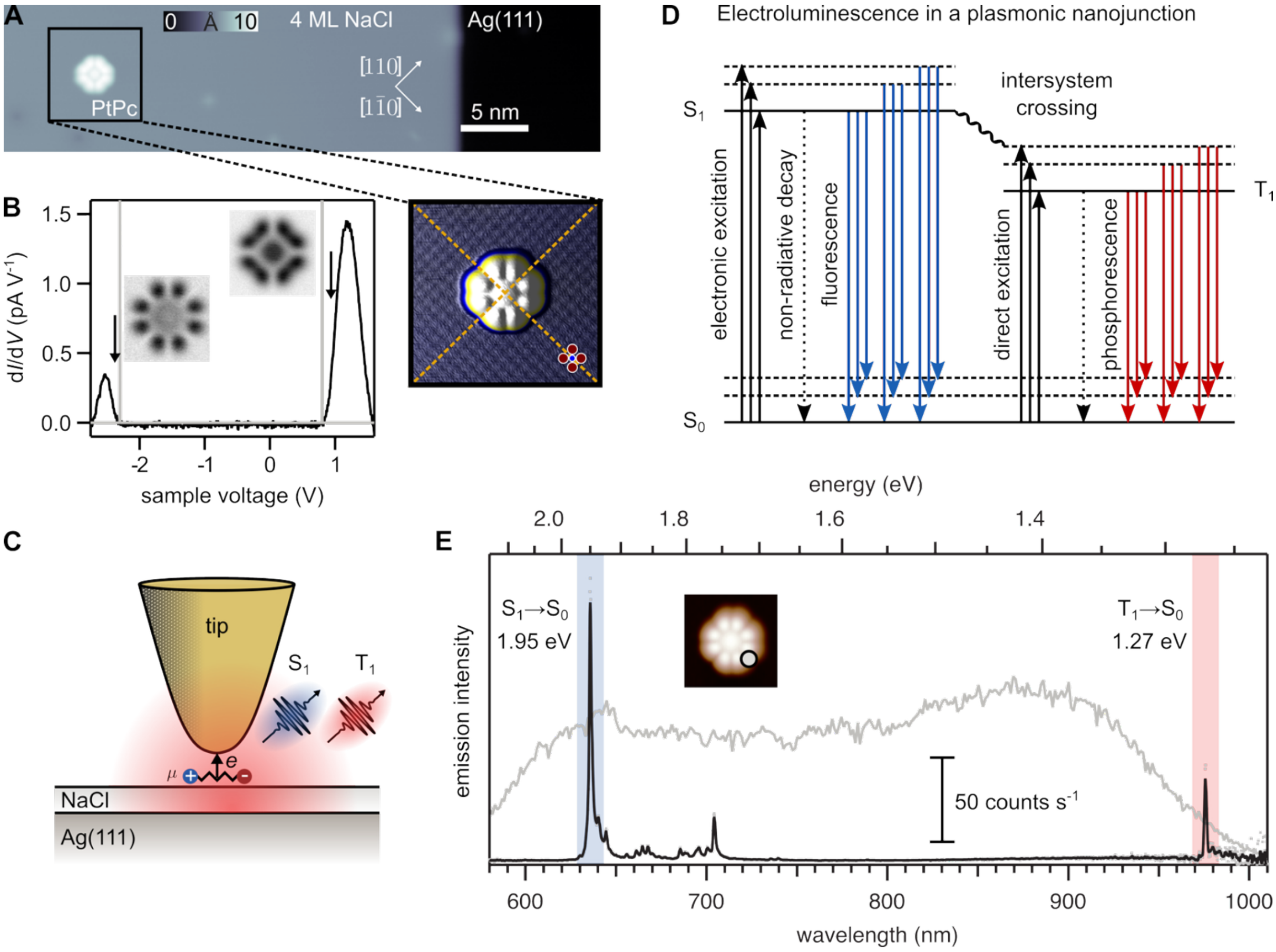

**Figure 1.** STML of a PtPc molecule. (A) Topography of a PtPc molecule adsorbed atop 4 ML NaCl/Ag(111) ($I$ = 3 pA, $V$ = 1 V). The inset shows a topograph ($I$ = 3pA, $V$ = -2.6V) with a color scale pronouncing the NaCl lattice corrugation overlaid with simulated illumination from the left-hand side, emphasizing the molecular features. (B) d$I$/d$V$ spectrum measured at the center of the molecule on 4ML NaCl with positive (HOMO) and negative (LUMO) ion resonances and onsets indicated by arrows at $V$ = -2.3 V and $V$ = 0.9 V, respectively. Gray lines mark the HOMO and LUMO energies computed by DFT. Right panels: Inverted contrast d$I$/d$V$ constant height maps at $V$ = -2.4 V (HOMO) and $V$ = 0.95 V (LUMO) (size: 3.5×3.5 nm$^2$, $V_{mod}$ = 50 mV). Arrows in the spectrum on the left mark the voltages at which these maps are recorded. (C) Schematic of the experiment with a single PtPc atop NaCl/Ag(111). Excitonic emission due to charge recombination in the molecule is detected outside of the UHV chamber. (D) Perrin-Jabłoński diagram of molecular emission processes in a tip-sample nanocavity with competing vibronic relaxation and radiation rates. (E) STML spectrum of a single PtPc molecule showing $S_1 \rightarrow S_0$ (blue) and $T_1 \rightarrow S_0$ (red) emission lines with their vibronic satellites (set point: $I$ = 80 pA, $V$ = -2.6 V, integration time $t$ = 60 s, grating: 300 grooves mm$^{-1}$). The spectrum has been normalized by a pure plasmonic spectrum, shown in gray ($I$ = 250 pA, $V$ = -2.6 V, $t$ = 1 s) recorded atop the NaCl layer at some distance from the molecule. The normalization accounts for both, spectral enhancement variations and wavelength-dependent detector sensitivity.

An STML spectrum for a single PtPc molecule, acquired at a negative sample voltage ($V$ = -2.6 V), with the tip placed on the edge of the PtPc molecule (gray dot in the inset), is shown in Fig. 1E. Sharp emission lines are observed at photon energies between 1.95 eV and 1.70 eV, and around 1.27 eV. The line at 1.95 eV is attributed to the electronic transition from the excited $S_1$ state to the $S_0$ ground state, as confirmed by prior STML[33, 45, 46] and photoluminescence measurements,[41] as well as time-dependent DFT (TD-DFT) calculations (see Supporting Table X3). Based on DFT calculations and Raman spectra of PtPc powder (see Supporting Fig. 3), we attribute the peaks at energies between 1.94 eV and 1.75 eV to vibronic satellites.

Previous studies of PtPc in single crystal form,[36] in solution,[37, 39] and in OLED devices[38] report phosphorescence energies between 1.28 eV and 1.31 eV. Accordingly, the sharp emission line at 1.27 eV is assigned to phosphorescence emission by radiative transition from the excited $T_1$ state to the $S_0$ ground state. Another emission line at 1.38 eV, rarely observed for negative sample voltages (see Supporting Fig. 4), coincides with a lifting of LUMO and LUMO+1 degeneracy, likely due to a defect in the NaCl lattice beneath the molecule. This emission line is attributed to luminescence from the cationic species $PtPc^+$, resembling the experimental conditions for cationic emission observed for free-base phthalocyanine.[47] No excitonic emission is observed under positive sample voltages, despite earlier reports of fluorescence of the anionic species at 1.36 eV in STML experiments on NaCl-covered Ag(100).[33]

We note that an earlier STML study of PtPc failed to observe the triplet emission line.[33] This discrepancy likely arises from the actual faintness of the triplet line, the requirement of efficient decoupling from the substrate (at least 3ML, ideally 4 ML of NaCl), and strong plasmonic enhancement preferably extending far into the red wavelength range. Our findings confirm the absence of the cationic emission, which we observe sporadically only on defected adsorption sites, as reported above. The allowed doublet emission $D_1^+ \rightarrow D_0^+$ of the cation requires a two-step excitation, in which the second step, $T_1 \rightarrow D_n^+$, appears to be not yet activated or at least inefficient at the applied voltages, as illustrated by the energy scheme of Fig. 2E. Alternatively, a direct excitation from the $S_0$ ground state to the $D_1^+$ state would be possible. However, that excitation path requires a bias voltage below -3.58 V (Fig. 2E) which is not reached in our experiments.

We performed TD-DFT calculations (see Supporting Information 1.2 and 1.3) for comparison with our experimental observations. These calculations yield $S_1$ and $T_1$ exciton energies of 2.02 eV and 1.03 eV, respectively, in fair agreement with the observed peaks at 1.95 eV and 1.27 eV. The calculated intensity ratio of the transition dipole moments $\mu_{T_1 \rightarrow S_0} : \mu_{S_1 \rightarrow S_0}$ is ~1:40 (see Supporting Information 1.3). Experimentally, we find that the $T_1$:$S_1$ intensity ratios range between ~1:3 (Fig. 1E) and ~1:10 (Fig. 4C) after correcting for the wavelength dependencies of the plasmonic enhancement and the detector sensitivity.

Next, we discuss the excited-state lifetimes τ derived from the observed linewidth using the uncertainty principle $\Delta E \Delta t \geq \hbar$). It is known that STML emission peaks may include multiple transitions and be broadened by multiple anti-Kasha transitions[12, 13].

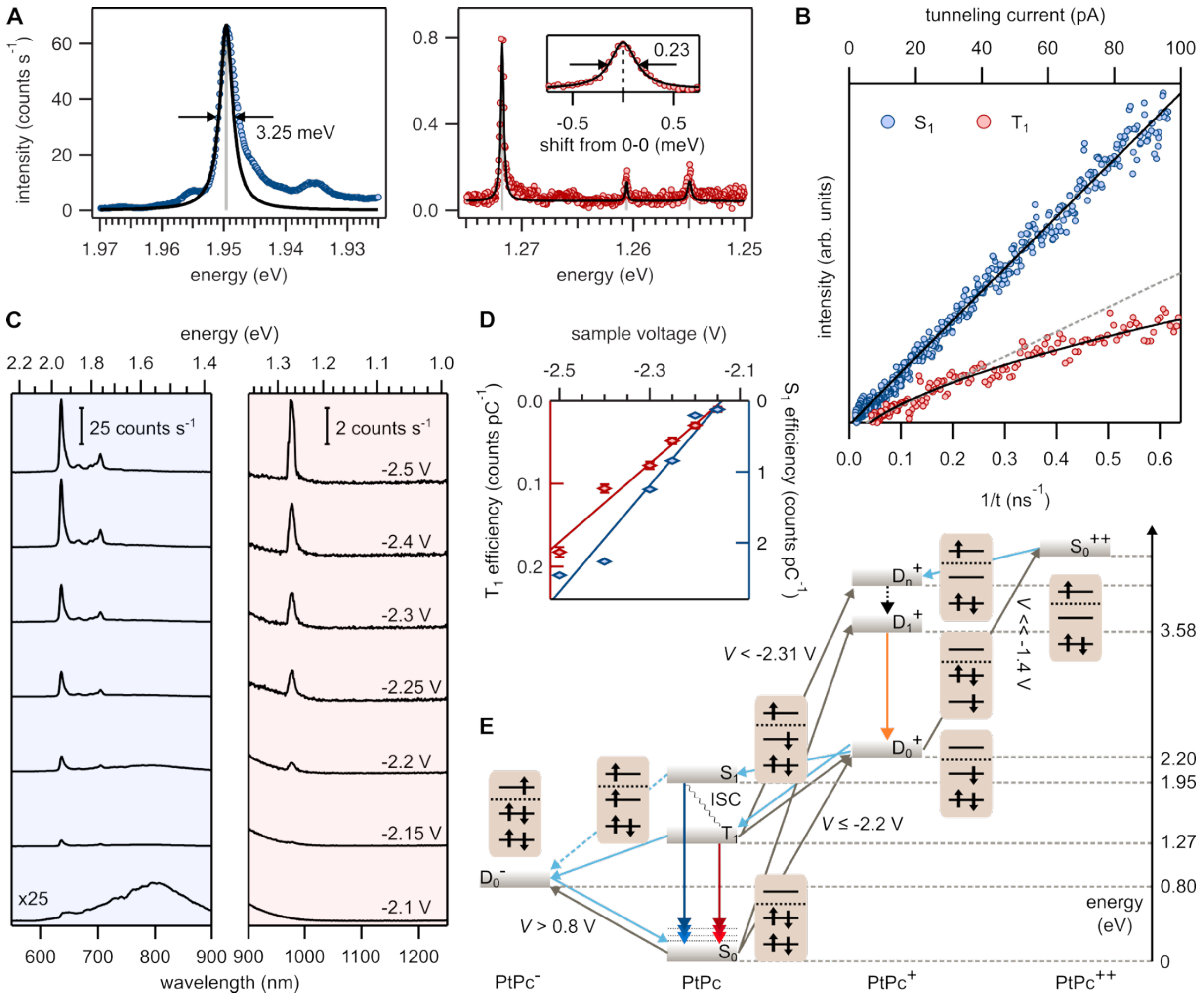

**Figure 2.** Electronic characterization of molecular emission. (A) Lorentzian line shape (black) fit of the $S_1$ (left) and $T_1$ (right) emission line and full-width at half-maximum FWHM. The gray lines mark the peak position of the Lorentzian. Inset (right). Blowup of the $T_1$ emission line with respect to peak position (1.272 eV) ($I$ = 80 pA, $V$ = - 2.6 V; for $S_1$: $t$ = 5 min; for $T_1$: $t$ = 50 min, grating: 1200 grooves mm$^{-1}$). (B) Current dependence of fluorescence and phosphorescence intensity at fixed voltage $V$ = -2.6 V (for $S_1$: $t$ = 200 ms, for $T_1$: $t$ = 1 s). The $S_1$ data is best fitted by $I^{0.99}$; the $T_1$ data is fitted using a three-state model.[63] The dashed gray line is a linear fit for the $T_1$ data in the range: 0 pA – 40 pA. (C) $S_1$ and $T_1$ spectra at the sample voltages indicated ($I$ = 200 pA, $t$ = 30 s, grating: 50 grooves mm$^{-1}$). The plasmonic spectrum (bottom) is measured at sample voltage $V$ = -2.1 V and is multiplied by 25 for clarity ($t$ = 1 s). (D) Integrated peak intensities from Lorentzian fits to the spectra in C. The linear extrapolation (solid lines) yields a cut-off voltage of $V$ = -2.15 V. (E) Energy diagram of accessible molecular states for 4 different charge states of PtPc.

Nevertheless, a longer $T_1$ lifetime would always result in sharper emission lines when controlling for other parameters. Figure 2A shows fluorescence (left) and phosphorescence (right) spectra obtained using a high-resolution 1200 grooves mm$^{-1}$ grating. A Lorentzian fit (black line in Fig. 2A) yields a linewidth of 2.49 meV ($\tau \geq 0.3$ ps) for fluorescence and 0.23 meV ($\tau \geq 2.9$ ps) for phosphorescence. We use only the high-energy side of the lines to determine the full linewidth because the strong peaks are asymmetrically broadened to the lower energy side. The evaluation thus suggests a substantially longer lifetime for the emission at 1.27 eV with respect to the 1.95 eV line, supporting our assignment of the 1.27 eV peak to phosphorescence. Note, however, that the life times given above only provide lower limits.

In order to reveal the exciton formation mechanism in PtPc, we analyze the voltage and current dependence of the luminescence intensity (see Figs. 2D and 2B, respectively). Both fluorescence and phosphorescence emissions onset simultaneously at a sample voltage of V ~ -2.15 V, which is well above the singlet emission energy of 1.95eV. The emission onset is thus not due to the quantum cut-off but coincides with the alignment of the tip Fermi energy with the edge of the positive ion resonance (-2.2 V, see Fig. 1B) within experimental error. This indicates that the first step in both $S_1$ and $T_1$ excitation is a removal of an electron from the HOMO. Although the existence of excitation by an inelastic process cannot be ruled out for less negative voltage, no emission is observed below the onset of the positive ion resonance.

We performed theoretical calculations based on the Pauli master equation approach combined with the DFT and TD-DFT calculations to simulate STML for the PtPc molecule adsorbed on NaCl films grown on Ag(111). These calculations reproduce the positive and negative ion resonances at sample voltages of +0.75 V and -2.35 V, respectively, corroborating the simultaneous onset of fluorescence and phosphorescence at the HOMO onset. These findings confirm that the excitation occurs by charge injection via the lowest-lying molecular cationic state.[48]

Next, we analyze the fluorescence and phosphorescence intensities at -2.6 V sample voltage as a function of tunneling current (Fig. 2B). We find a linear onset for both emission intensities, confirming a one-electron tunneling process via hole creation in the HOMO, as discussed above. An electron capture from the metal substrate by the LUMO is the second step, necessary to create the exciton, as illustrated in Fig. 2E. This second charge transfer with the substrate restores molecular charge neutrality and thus preserves linearity in the tunneling current. However, at currents exceeding 80 pA, the phosphorescence intensity starts to deviate from linearity. At this current the time between tunneling electrons is still as low as ~2 ns. This behavior can be attributed to the onset of saturation of triplet occupancy, allowing us to estimate an upper limit for the triplet lifetime to be ~1 ns. The measured line width, however, indicates a lower bound for the triplet lifetime limit of ~3 ps (as discussed above). It is worth noting that the early saturation of $T_1$ state may be enhanced by an increased non-radiative decay rate as the metallic tip approaches the molecule with increasing tunneling current. Compared to the $S_1$ state, the $T_1$ state will generally be more susceptible to variations in non-radiative decay due to its inherently longer lifetime.

Figure 2E provides an overview of the energy levels of PtPc for the excitation and emission of the $S_1$ and $T_1$ states. Experimentally accessible energies are marked on the right-hand scale. As established earlier, the first step is the extraction of an electron from the HOMO, leading to the formation of a doublet state of the cationic species, $D_0^+$. Depending on the spin of the injected electron, this facilitates population of either the $S_1$ or $T_1$ state. When excitons are generated via charge-injection, a simple spin-statistics argument suggests a 3:1 ratio favoring $T_1$ over $S_1$ excitation due to the multiplicities of these states. [22] However, this ratio may be slightly reduced because the tunneling barrier for triplet excitation is higher than that for singlet excitation.[34]

A longer lifetime of the triplet state relative to the singlet state further implies that the triplet state occupation will be on average higher than the singlet state occupation. Nevertheless, the contrasting experimentally observed low phosphorescence to fluorescence intensity ratio, ranging from ~1:3 to ~1:10 (see Figs. 1E, 2C, 4C) can be explained by two other factors. First, the coupling strengths for radiative decay to the plasmon field likely differs between the two types of emission, despite both fluorescence and phosphorescence being enhanced by the localized plasmon field[49-51] .To account for the pure wavelength dependence of this enhancement and detector sensitivity, the spectra have already been divided by the tip plasmonic spectrum. Second, the triplet state lifetime is long enough to let non-radiative decay channels dominate the decay. The possible decay of the triplet state $T_1$ to the anionic $D_0^-$ and especially to the cationic $D_0^+$ state (see Fig. 2E) through an "exciton ionization"-like process within the STM electric field will be very important and may best explain the absence of strong phosphorescence. This decay path may also explain why we do not observe a triplet life time in the range of 10 - 100 μs, as reported for the pentacene molecule by Peng et al.[52] or 170 - 670 μs reported for PTCDA by Sellies *et al.*[32]. In these experiments, a spontaneous decay to a charged state is prohibited. In addition, the theoretical exciton decay rates obtained for the gas phase (see Methods) cannot serve as a realistic comparison for our experiment because the calculation assumes a perfect decoupling from the metallic substrate. We note that, under similar experimental conditions (few ML NaCl atop a metal substrate), Kaiser et al.[34] derived triplet state lifetimes for ZnPc between 0.09 ns and 0.37 ns using a dynamic model.

To further assess the role of the cationic molecular state in emission, we return to the energy diagram in Fig. 2E. The $D_1^+ \rightarrow D_0^+$ transition requires a two-step charge injection process. The first step involves charge injection from the ground state $S_0$ of the neutral molecule to $D_0^+$, which is reached at the HOMO onset energy. From $D_0^+$, a second excitation step to $S_0^{++}$ can relax to the excited state $D_1^+$. This intermediate doubly charged state must overcome a strong intra-molecular Coulomb repulsion. Alternatively, the $D_0^+$ state may first relax to the $S_1$ or $T_1$ states, from where a second charge injection step is needed to reach $D_n^+$. The second excitation step thus starts either from the $S_1$ state, which decays efficiently by fluorescence, or from the $T_1$ state, which can readily decay into the $D_0^-$ state. Although we cannot determine the precise energy of the $D_n^+$ state in Fig. 2E, we may assume that the bias voltage necessary to reach it from $T_1$ should lie below -2.2V thus introducing a second threshold voltage close to the threshold of $D_0^+$ creation. Note that a direct excitation from $T_1 \rightarrow D_1^+$ is not possible in a single injection step. Most importantly, both of these routes would have a clear signature in the form of a super-linear tunneling current dependence. In contrast, the data in Fig. 2D do not show such behavior. Based on this evidence and the observation of the 1.27 eV emission line in photoluminescence (see next section), we conclude that this emission cannot originate from a charged trionic state.

### STM tip-enhanced photoluminescence (TEPL) of a single phosphor

Besides the direct excitation of the $T_1$ state by charge injection, ISC from the $S_1$ to the $T_1$ state also contributes to phosphorescence.[22, 42, 53] To investigate the role of ISC in single molecule phosphorescence in a controlled manner, we study PtPc using the recently demonstrated method of resonant TEPL[12-14] with laser light resonantly exciting the $S_1$ state, thereby maximizing molecular excitation. Figure 3A illustrates the experimental setup. A low-pass filter (LPF) blocks the excitation line, while a charge-coupled device (CCD) camera detects the phosphorescence. For excitation, the laser is tuned to the $S_1$ absorption line to ensure that the $T_1$ state is populated predominantly via ISC from the $S_1$ state, with no contribution from indirect excitation of the $T_1$ state via charged states as in STML (see Perrin-Jabłoński diagram in Fig. 3B).

The TEPL data (Fig. 3D) is recorded at a low tunneling current ($I$ = 3 pA) and a sample voltage of $V$ = 1 V which excludes any plasmon-mediated optical excitation of the molecule. Moreover, the STM tip is positioned at a lateral distance of 1.9 nm from the molecule's center (as indicated by the grey ring in Fig. 3C) to exclude any electron injection into the molecule. The observed STM-PL spectrum reveals vibronic satellites of the $S_1$ peaks down to 1.6 eV and $T_1$ emission at 1.264 eV, in close agreement with the STML spectrum in Fig. 1E (see also Supporting Fig. 3). To the best of our knowledge, this is the first direct observation of ISC in a single-molecule STM experiment. A slight shift (~8 meV) in phosphorescence between STML (1.272 eV, $V$ = -2.6 V) and TEPL (1.264 eV, $V$ = 1 V) is observed, which can be attributed to tip-induced effects, like the Stark shift as the measurements are performed with different applied voltages [9, 27, 54, 55] (see schematic in Fig. 4A). Additional measurements of the Stark effect in $T_1$ emission are shown in Supporting Figures 6 and 7.

The linewidth of the phosphorescence in TEPL (0.21 meV, inset in Fig. 3D) is comparable to that in STML (0.23 meV, right panel in Fig. 2A), suggesting that the lifetime of the $T_1$ state does not differ significantly between the two excitation methods. Combined with the saturation of phosphorescence with increasing tunneling current (Fig. 2B), this finding affirms that the lifetime of $T_1$ state is comparable to or longer than the vibrational relaxation time.

It is important to note that the intensity ratio between phosphorescence and fluorescence is not directly accessible in resonant TEPL. Since the $S_0 \rightarrow S_1$ transition is used for excitation, a strong background intensity is present on the fluorescence line, which must be suppressed using an LPF. Thus, we use the $S_1$ vibronic satellites present in both STML (Fig. 1D) and TEPL (Fig. 3D) to compare the phosphorescence intensities for both excitation methods. The ratio of phosphorescence intensity to the major $S_1$ vibronic peaks is approximately two orders of magnitude lower in TEPL than in STML. The phosphorescence intensity observed in TEPL directly provides a lower bound for the ISC contribution in STML and constitutes a clear observation of the occurrence of ISC in an STM experiment at the single-molecule level.

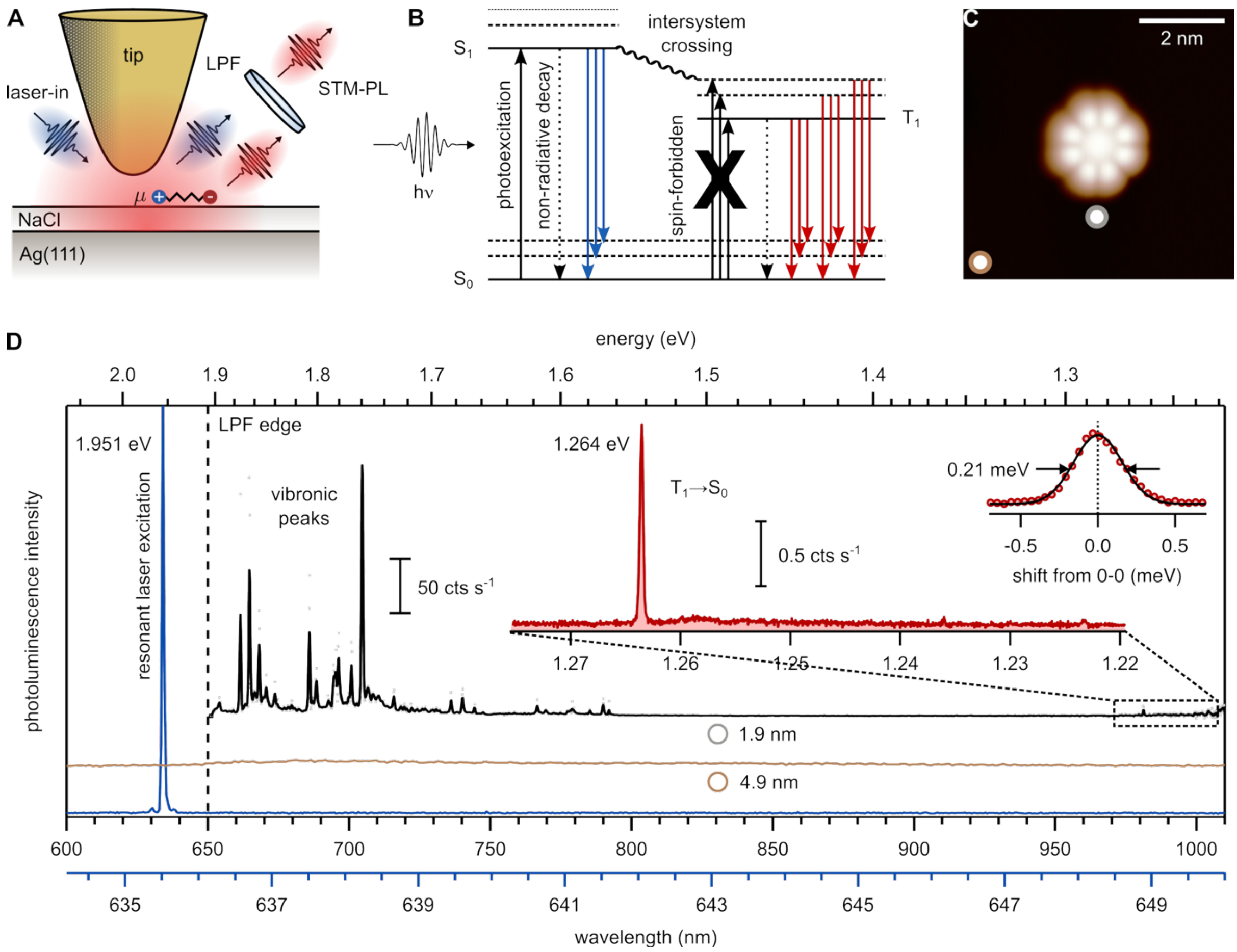

**Figure 3.** $T_1$ emission in TEPL. (A) Schematic illustration of the TEPL experiment in which $S_1$ is resonantly excited by incident laser light. (B) Schematic demonstrating that $T_1$ can only be excited via intersystem crossing (ISC). (C) STM topography image of the PtPc molecule adsorbed atop 4 ML NaCl on Ag(111) ($I$ = 3 pA, $V$ = -2.6 V). (D) Laser line (blue, 635.52 nm, grating: 1800 grooves $mm^{-1}$) used for resonant photoexcitation of the $S_0 \rightarrow S_1$ transition. The laser spectrum is plotted with respect to the lower of the two wavelength scales. TEPL spectra measured at the positions indicated by the white dots with colored rings in (C). Distance from the center of the molecule: brown, $r$ = 4.9 nm; gray, $r$ = 1.9 nm (laser power: 1 μW, $t$ = 60 s, grating: 300 grooves $mm^{-1}$). The dashed line marks the 650 nm edge of the long-pass filter LPF. Inset: Zoom-in on the phosphorescence region of the TEPL spectrum ($I$ = 3 pA, $V$ = 1 V; laser power: 1 μW, $t$ = 60 min, grating: 1200 grooves $mm^{-1}$). The second inset shows the detailed $T_1 \rightarrow S_0$ transition and a Lorentzian fit of the $T_1$ line with the indicated FWHM.

## Intersystem crossing and exciton-plasmon coupling

In order to estimate the ISC rate, we analyzed the STML and TEPL spectra under the assumption that the intramolecular dynamics is similar in both experiments. By comparing the $T_1$:$S_1$ intensity ratios, we find that the ISC rate is about one tenth of the singlet decay rate. For the detailed calculation and the approximations used, see the Supplementary Information 1.4 and Supporting Figure 1. Based on the observed fluorescence line width (as discussed above), the ISC rate is coarsely estimated at ~0.3 $ps^{-1}$. This value aligns well with literature value (0.67 $ps^{-1}$) of Pd-Phthalocyanine in α-chloronaphtalene at 4 K. [37]

While assessing the ISC rate required the assumption that electroluminescence and photoluminescence follow similar intramolecular dynamics, detailed investigations reveal deviations when the tip position relative to the molecule varies. In order to quantify the effects induced by plasmonic coupling, we investigate spectra recorded at varying tip position. An overview of the angular dependence of STML and TEPL data are presented in Supporting Figure 5.

First, we record constant current STML spectra with the tip placed at different positions on a circular path around the molecule's center (Fig. 4B) at a radius $r$ = 1.1 nm and azimuths $\theta = n \cdot 45°$, where $n$ is an integer. The results indicate that the $T_1$:$S_1$ intensity ratio is significantly higher at $\theta$ = 45° (Fig. 4C), demonstrating the critical role of the position of the plasmonic tip. The charge injection process, which is similar for the generation of both excitons, would not explain a dependence on θ. Figure 4D shows that the higher $T_1$:$S_1$ intensity ratio is generally higher at azimuthal angles $\theta = (2n + 1) \cdot 45°$ and lower at $\theta = n \cdot 90°$. The ratio is in anti-phase with the linewidths of both fluorescence and phosphorescence (Figs. 4E, 4F), which broaden with increased X-P coupling strength. [2, 9, 40, 56, 57]

Specifically, the $S_1$ linewidth variation of ~22% (Fig. 4E) indicates a shortening of the $S_1$ lifetime at $\theta = n \cdot 90°$, while the $T_1$ linewidth (Fig. 4F) variation of ~29% is in-phase with the $S_1$ linewidth broadening. In contrast to the X-P coupling, it is reasonable to assume that the ISC rate is not significantly affected by the tip position, as ISC is primarily induced by the Pt atom at the molecular center[42] rather than by the atoms of the tip. The lifetime shortening observed is therefore mainly attributed to enhanced X-P coupling, which accelerates the radiative rate from $S_1$, thereby reducing the branching ratio of the competing ISC. Since ISC converts $S_1$ to $T_1$, it directly alters the $T_1$:$S_1$ intensity ratio. Therefore, the variation observed in Fig. 4D is considered to reflect changes in the ISC branching ratio induced by changes in X-P coupling. We remark that a concomitant change of the radiative-to-non-radiative branching ratio of the $T_1$ state itself could also affect a detailed quantitative evaluation. The tip position dependency of the $T_1$:$S_1$ ratio provides a means to control the weight between current-induced fluorescence and phosphorescence, thereby enabling emission color tuning in plasmonic nanocavities.

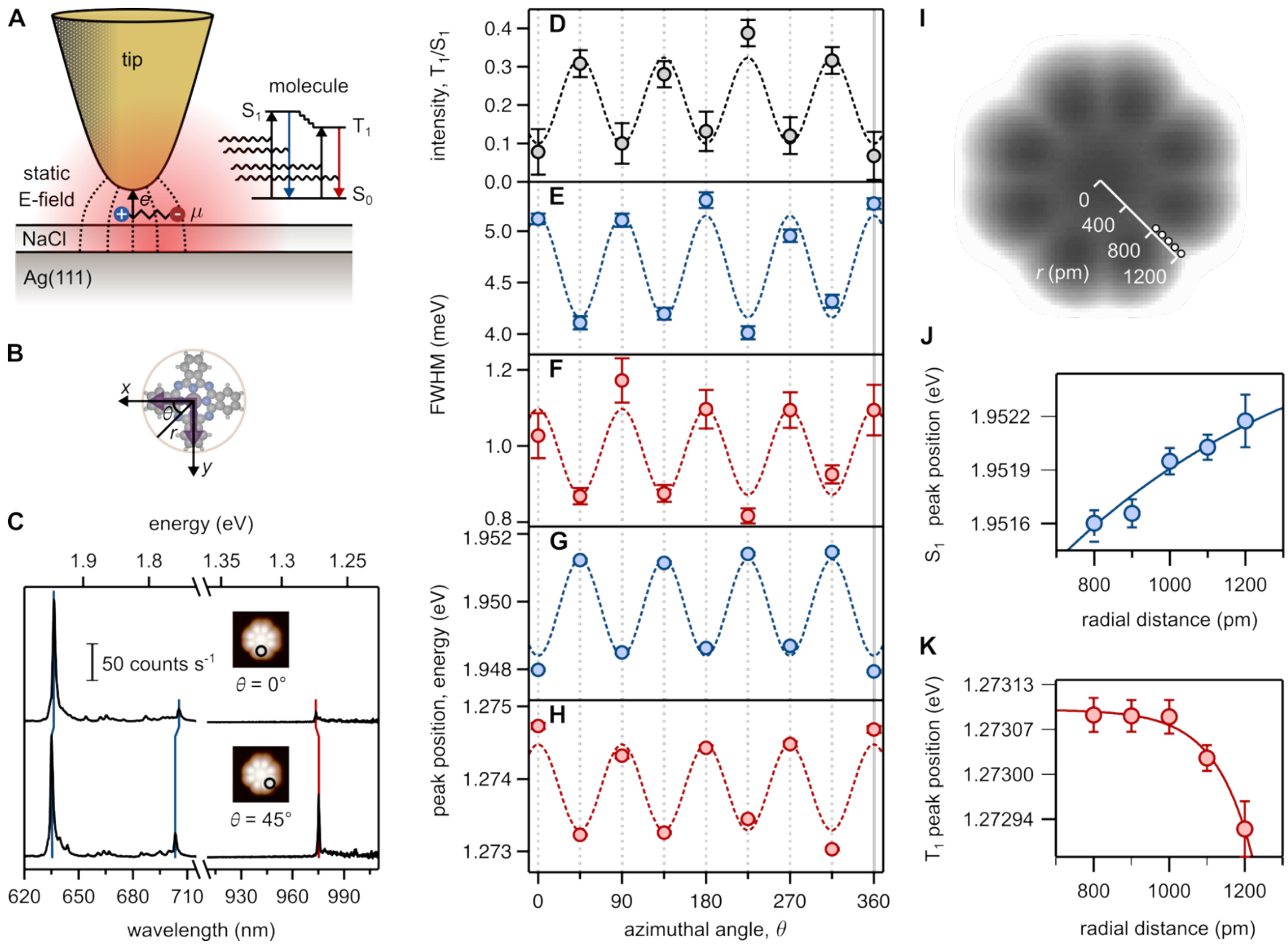

**Figure 4.** Azimuthal and distance dependence of $S_1$ and $T_1$ emission. (A) Illustration of the interaction between the plasmonic field in the nanocavity and the molecular fluorescence and phosphorescence together with the tip-induced electric field gradient over the molecule (red). (B) Schematic defining the polar coordinates r and θ with respect to the molecular geometry. (C) STML spectra obtained for θ = 0° and 45° ($r$ = 1.1 nm; $I$ = 50 pA, $V$ = -2.6 V, for $S_1$: $t$ = 5 s and for $T_1$: $t$ = 30 s; grating: 300 grooves mm$^{-1}$). (D) Azimuthal dependence of the $T_1$ intensity normalized to the $S_1$ intensity. The intensities are obtained as peak areas under Lorentzian fits. Error bars represent standard deviation of the peak area in the fit. (E, F) Azimuthal dependence of $S_1$ (blue) and $T_1$ (red) peak widths. (G, H) Azimuthal dependence of peak position of $S_1$ and $T_1$ emission lines obtained from Lorentzian fits. (I) STM topography image of a PtPc molecule with tip distance scale ($I$ = 4 pA, $V$ = -2.6 V). (J, K) $S_1$ and $T_1$ peak energies as a function of radial distance from the center of the molecule. The solid lines are exponential fits to guide the eye. Intensity ratio and line widths as function of the radial distance for this data series are shown in Supporting Figure 8.

In Fig. 4G, the $S_1$ peak energy at azimuthal positions θ = (2n+1) · 45° is blueshifted compared to $\theta$ = n · 90°. This observation is consistent with earlier studies reporting Lamb effect-induced redshift of the $S_1$ emission line.[9, 11, 12, 57] However, we find that the energy shift of the phosphorescence line as a function of azimuth (Fig. 4H) is opposite to the one of the fluorescence line. A similar opposite behavior between fluorescence (Fig. 4J) and phosphorescence (Fig. 4K) is observed when the tip is moved away from the center of the molecule (Fig. 4I). These suggest competing Lamb and Stark effects, acting in opposite directions but with different amplitudes for the two transitions. Such behavior has been reported previously,[9] but a quantitative disentanglement of the contributions would require further investigation.

## CONCLUSIONS

This study provides a detailed picture of nanoscale processes in plasmon-enhanced phosphorescence. The unique combination of STML and TEPL facilitates the analysis of phosphorescence and allows a direct observation of singlet-triplet ISC of excitons at the single-molecule scale. The study demonstrates that coupling to a plasmonic cavity significantly affects the intensities of the singlet and triplet excitons of an isolated PtPc molecule. For the used NaCl/Ag(111) substrate, both excitons are observed in STML only under negative polarity, with their onset at the alignment of the HOMO with the tip Fermi energy. The findings indicates that phosphorescence – when corrected for plasmonic enhancement by the STM tip – even though observable, remains much weaker than fluorescence due to more efficient non-radiative decay pathways for the triplet compared to the singlet. The coupling to a plasmonic structure shortens the singlet lifetime, increases fluorescence intensity, reduces triplet occupation via ISC, and consequently results in lower phosphorescence intensity. These results suggest that X-P coupling could be used to modulate fluorescence and phosphorescence in OLED applications.[27] Future studies could explore phosphors whose triplet and singlet transitions are closer to each other in energy, allowing for an increased ISC. Moreover, suppressing the decay of the triplet state to anionic or cationic molecular states may significantly improve phosphorescence efficiency. The experimental insights presented here have implications for plasmonic OLED devices,[27, 58] bioimaging,[59, 60] and biosensing[61, 62], where exciton dynamics play a central role.

## AUTHOR CONTRIBUTIONS

A.G., H.I., M.I.-I., K.Ku., K.Ke., and Y.K. conceived and designed the research program. H.I., M.I.-I., and Y.K. constructed the experimental set-up. A.G., M.I.-I., H.I., K.Ku., and R.B.J. performed the experiment using STM and analyzed the data. K.M. conducted the theoretical calculations. Y.K. and K.Ke. directed the project. All authors discussed the results and wrote the manuscript.

## FUNDING

This work was financially supported in part by Japan Society for the Promotion of Science (JSPS) KAKENHI: Grant Nos. 22H04967 (Y. K.), 21H04644 (H.I.), 21K14481 (K.M.), 21H05412 (K.K.), 21K14593 (K.K.), 22K18959 (M.I.-I.) and Japan Science and Technology Agency (JST) PRESTO: Grant No. JPMJPR22A2 (M.I.-I.).

## ACKNOWLEDGEMENTS

A.G. acknowledges all members of the Surface and Interface Science Laboratory, RIKEN for their hospitality during the research stay. A.G. and K.Ku. would like to thank Olle Gunnarsson, Christopher C. Leon, and Anna Rosławska for stimulating discussions. Parts of numerical calculations were performed using the HOKUSAI system at RIKEN and the Research Center for Computational Science (RCCS) system at Institute for Molecular Science (IMS).

## ABBREVIATIONS

STM, scanning tunneling microscopy; UCEL, electroluminescence upconversion; ISC: intersystem crossing; SOC : spin-orbit coupling ; OLED : organic light emitting diode; PtPc: platinum (II) phthalocyanine ; UHV: ultra-high vacuum; ML: monolayer; X-P interaction: exciton-plasmon interaction; HOMO: highest occupied molecular orbital; LUMO: lowest unoccupied molecular orbital; STML: STM-induced electroluminescence; TEPL: tip-enhanced photoluminescence; DFT: density functional theory; TD-DFT: time-dependent density functional theory; LPF: low-pass filter; CCD: charge-coupled device.

# 1. Supporting information

## 1.1 Sample preparation and details of STM measurements

The Ag(111) single-crystal (> 99.999% purity) is cleaned by repeated cycles of $Ar^+$ ion sputtering and subsequent annealing. NaCl is then evaporated thermally from a Knudsen cell heated to 890 K on the Ag(111) surface held at room temperature. For these preparation conditions, defect-free (100)-terminated two, three, and four monolayers thick NaCl islands are obtained. PtPc is deposited onto the NaCl-covered Ag(111) directly in the STM head at 4.5 K - 10 K using a homemade evaporator heated to 638 K. An electrochemically etched gold wire[1] (99.95% purity) is used as tip and conditioned by controlled indentation and voltage pulses on the clean Ag(111) surface. For the experiment, it is crucial to have broad plasmonic resonances spanning the range from $S_1$ to $T_1$ emission of PtPc. Therefore, the plasmonic resonance of the tip-sample junction is characterized using STML. A typical resonance of the tip-sample cavity plasmon is shown in Fig. 1E (grey trace).

## 1.2 DFT and TD-DFT analysis of the molecular luminescence spectra including vibronic transitions and transition energies between molecular many-body states

Electronic and vibrational structures of PtPc were analyzed using first-principles calculations based on density functional theory (DFT) and time-dependent DFT (TD-DFT) implemented in the software package Gaussian 16.[2] All calculations were performed using the Ahlrichs triple-zeta valence basis set with polarization and diffuse functions (def2-TZVPD ).[3] Effective core potential was utilized to substitute the 60 core-orbitals of the Pt atom.[4] First, the geometry for the ground electronic state of the neutral molecule was optimized using the hybrid B3LYP functional.[5] Next, according to the tuning procedures,[6] the optimal value of the system-specific range separation parameter ▯ was determined as 0.151 $Bohr^{-1}$ for a range-separated hybrid density functional (LC- wHPBE).[7] Hereafter, the tuned range-separated functional was referred as LC-wHPBE*. All subsequent calculations were performed at the LC-wHPBE*/def2-TZVPD level of theory.

The geometry optimizations were carried out for the ground electronic state of different charged molecules. The analysis of the vibrational frequencies for each charged molecule was performed to ensure that all positive frequencies were obtained. The simulations yield total energies $E_{N,a}$ of electronic many-body state $|N,a\rangle$, where $N$ indicates the number of excess electrons with respect to the neutral molecule. The simulation results were shown in Supplementary Table X1. The results indicate the cationic, neutral, and anionic states ($N = -1, 0, 1$) can be accessed in the range of bias voltage used in the experiment. We thus focused on these charged states and excluded the doubly charged states ($N = \pm 2$).

**Supporting Table X1** | The number of excess electrons $N$ with respect to the neutral molecule, spin multiplicity $s$, total energy $E_{N,a}$ (sum of the electronic and zero-point vibrational energies), and energy difference from the total energy of $|0, S_0\rangle$.

| Electronic state | $N$ | $s$ | $E_{N,a}$ (eV) | $\Delta E_{N,a}$ (eV) |
|---|---|---|---|---|
| $\|0, S_0\rangle$ | 0 | singlet | −48579.831781 | 0 |
| $\|-1, D_0\rangle$ | −1 | doublet | −48573.441673 | +6.390107 |
| $\|-2, S_0\rangle$ | −2 | singlet | −48563.839070 | +15.992711 |
| $\|-2, T_0\rangle$ | −2 | triplet | −48563.018210 | +16.813570 |
| $\|+1, D_0\rangle$ | +1 | doublet | −48582.017591 | −2.185810 |
| $\|+2, T_0\rangle$ | +2 | triplet | −48581.101356 | −1.269575 |
| $\|+2, S_0\rangle$ | +2 | singlet | −48581.119451 | −1.287670 |

The vertical excitation energies for the neutral, cationic, and anionic states were calculated using the TDDFT at the LC-wHPBE*/ Def2TZVPPD level. Supporting Table X2 exhibits the corresponding energies and transition dipole moments. To analyze molecular luminescence spectra, geometry optimization and frequency analysis for the first excited electronic states with both singlet and triplet spin multiplicities were performed. The simulation results of the total energies were displayed in Supporting Table X3. The obtained information on the molecular vibrations was utilized to evaluate the vibrational overlap integrals $\langle v_{n'}^{N,a'} | v_n^{N,a} \rangle$ associated with the optical transition of the neutral molecules, using the method implemented in Gaussian 16.[8, 9] The results are shown in Supporting Figure 2.

Image interaction energy for PtPc molecule adsorbed on a NaCl ultrathin film growth on a metal surface was evaluated using the dielectric model introduced by Barone *et al.*.[9] The electron and hole attachment energies for the adsorbed molecule are shifted from the values for a molecule in the gas phase, because the molecule is positioned near the metal substrate. It had been reported that the differences are mainly attributed to the image interaction with the metal substrate. Following the literature, the shift for the electron/hole attachment energies was estimated as 0.769 eV. Transition energy $\varepsilon_{Na,Mb}$between molecular many-body states $|N,a\rangle$ and $|M,b\rangle$ is calculated as $\varepsilon_{Na,Mb} = E_{N,a} + E_{N,a}^{\text{Imag}} - E_{M,b} - E_{M,b}^{\text{Imag}}$, where $E_{N,a}^{\text{Imag}}$ indicates the correction of the total energies of the molecule owing to the image charge effects. Calculation results were summarized in Supporting Table X4. As the work function for NaCl/Ag (111) can be estimated as 3.57 eV,[10] we consider the following molecular many-body states in the theoretical analysis: $|0, S_0\rangle$, $|0, S_1\rangle$, $|0, S_2\rangle$, $|0, T_1^{m=0,\pm1}\rangle$, $|0, T_2^{m=0,\pm1}\rangle$, $|-1, D_0^{\sigma=\pm1/2}\rangle$, $|+1, D_0^{\sigma=\pm1/2}\rangle$, and $|+1, D_1^{\sigma=\pm1/2}\rangle$. It is noteworthy that $|0, S_1\rangle$, $|0, T_1^m\rangle$, and $|+1, D_0^\sigma\rangle$ are, respectively, energetically degenerated with $|0, S_2\rangle$, $|0, T_2^m\rangle$, and $|+1, D_1^\sigma\rangle$.

Supporting Table X2 | Vertical excitation energies Eex from the ground electronic state to several lowest excited states and transition dipole moment $\mu$ for the neutral ($N = 0$), cationic ($N = -1$), and anionic states ($N = +1$) of the molecule. For the neutral molecule, the singlet spin multiplicity was accounted.

| | $N = 0$ | | $N = -1$ | | $N = +1$ | |
|---|---|---|---|---|---|---|
| | $E_{ex}$ (eV) | $\mu$ (a.u.) | $E_{ex}$ (eV) | $\mu$ (a.u.) | $E_{ex}$ (eV) | $\mu$ (a.u.) |
| 1 | 2.1591 | 2.7950 | 1.1699 | 0.3493 | 0.3532 | 0.0000 |
| 2 | 2.1591 | 2.7950 | 1.1699 | 0.3499 | 1.4666 | 0.6077 |
| 3 | 2.6076 | 0.0000 | 1.6838 | 0.0000 | 1.5546 | 0.0407 |
| 4 | 2.6985 | 0.0000 | 1.8407 | 1.2856 | 2.0013 | 1.0420 |
| 5 | 2.7931 | 0.0000 | 1.8407 | 1.2866 | 2.0629 | 2.9645 |

Supporting Table X3 | Total energy $E_{N,a}$ (sum of the electronic and zero-point vibrational energies) for singlet and triplet excited electronic states and energy difference from the total energy of $|0, S_0\rangle$.

| Electronic state | $E_{N,a}$ (eV) | $\Delta E_{N,a}$ (eV) |
|---|---|---|
| $|0, S_1\rangle$ | −48577.571532 | 2.019167 |
| $|0, T_1\rangle$ | −48578.557700 | 1.032999 |

Supporting Table X4 | Transition energy $\varepsilon_{Na,Mb}$ between molecular many-body states $|N,a\rangle$ and $|M,b\rangle$.

| $\|N,a\rangle$ | $\|M,b\rangle$ | $\varepsilon_{Na,Mb}$ (eV) |
|---|---|---|
| $\|0,S_0\rangle$ | $\|-1,D_0^{\sigma}\rangle$ | −5.6221 |
| $\|0,S_1\rangle$ | $\|-1,D_0^{\sigma}\rangle$ | −3.6029 |
| $\|0,T_1^{m}\rangle$ | $\|-1,D_0^{\sigma}\rangle$ | −4.5891 |
| $\|+1,D_0^{\sigma}\rangle$ | $\|0,S_0\rangle$ | −2.9549 |
| $\|+1,D_0^{\sigma}\rangle$ | $\|0,S_1\rangle$ | −4.9740 |
| $\|+1,D_0^{\sigma}\rangle$ | $\|0,T_1^{m}\rangle$ | −3.9879 |

### 1.3 DFT and TD-DFT analysis of the optical transition rates for the gas phase molecule

Optical transition rates were evaluated with the DFT and TDDFT calculations implemented in the ORCA software package.[11] As the spin-orbit coupling (SOC) should be included to evaluate the phosphorescence and ISC rates, we utilized this software package. Molecular structure for the first excited electronic state with the singlet spin multiplicity ($S_1$ state) was optimized at the B3LYP/def2-TZVP level of theory. To accelerate the calculation, the resolution of identity approximation for the Coulomb part (RIJ) and the chain of sphere algorithm for the exchange part (COSX) were employed with the corresponding auxiliary basis sets.[12]

The relativistic effects were included by employing the zeroth-order regular approximation (ZORA) method.[13] The B3LYP functional was utilized with the scalar relativistic (SR) contracted version of the Ahlrichs triple-zeta valence basis set with polarization functions (ZORA-def2-TZVP) for H, C, and N atoms, and the segmented all-electron relativistically contracted (SARC-ZORA-TZVP) basis set for Pt atom.[2] The RIJCOSX approximation with auxiliary basis sets (SARC/J and def2-TZVP/C) was utilized to accelerate the calculations.[12, 14] Molecular geometry with the $T_1$ state was optimized with the SR ZORA Hamiltonian. Then, at the optimized geometry, the SOC was included as a perturbation into the SR ZORA calculation results, where 25 singlet and 25 triplet excited states were used as the basis for the perturbation expansion, to find the spin-mixed states and the finite amount of the transition dipole moments for the triplet-origin states.[15] According to the previous study,[16] the spin-orbit integrals were calculated with the RI-

SOMF(1X) approximation[17] and TDDFT calculations were performed without Tamm-Dancoff approximation (TDA).[18]

The transition dipole moment $\mu_{S_0-S_1}$ for the $S_0 - S_1$ transition was evaluated at the optimized structure for the $S_1$ state and was obtained as 2.93 (a.u.). For the $S_0 - T_1$ transition, $\mu_{S_0-T_1}$ = 0.0686 (a.u.) was obtained at the optimized structure for the $T_1$ state. The fluorescence and phosphorescence rates were estimated using the formula $k = \frac{4}{3\hbar}\left(\frac{\omega}{c}\right)^3 |\mu|^2$ with $\hbar\omega$ being the optical transition energy, $\hbar$ the Planck constant divided by $2\pi$, $c$ the speed of light in vacuum, and $\mu$ the transition dipole moment.[19] The rate $k_{S_0-S_1} = 5 \times 10^7$ s$^{-1}$ and $k_{S_0-T_1} = 5 \times 10^3$ s$^{-1}$ were, respectively, utilized for the fluorescence and phosphorescence processes in the gas phase, that are at the same order of magnitude as calculated with the above-shown values of $\mu_{S_0-S_1}$ and $\mu_{S_0-T_1}$.

### 1.4 Estimation of the intersystem crossing rate

Supporting Figure 1 shows an oversimplified energy scheme (compare to Fig.2E) for electroluminescence of the studied PtPc molecule. In order to explore the steady state condition, we set up master equations for the singlet ($n_s$) and triplet ($n_T$) occupation numbers as follows:

$$\dot{n}_S = I_{el}\,\beta - n_S \frac{\tau'_S + \tau_{ISC}}{\tau'_S\ \tau_{ISC}} \xrightarrow{equilib} 0$$
$$\dot{n}_T = I_{el}(1-\beta) + n_S \frac{1}{\tau_{ISC}} - n_T \frac{1}{\tau'_T} \xrightarrow{equilib} 0$$

**(eqns. 1)**

Here $\tau_{ISC}$ is the inverse intersystem crossing rate, $\tau'_S$ the effective singlet state lifetime (composed of non-radiative $\tau_{nrS}$ and radiative $\tau_{rS}$ singlet lifetimes) and $\tau'_T$ the effective triplet state lifetime (see also Supporting Figure 1). Eqns. 1 describe the excitation by a tunnel current $I_{el}$. The spin multiplicity of triplet and singlet excitation is assumed to be 3:1 providing an excitation branching of β = 0.25 to the singlet state and 1 - β = 0.75 to the triplet state. Both time derivatives tend towards zero when equilibrium is reached.

$$\dot{n}_S = I_{photo} - n_S \frac{\tau'_S + \tau_{ISC}}{\tau'_S\ \tau_{ISC}} \xrightarrow{equilib} 0$$
$$\dot{n}_T = n_S \frac{1}{\tau_{ISC}} - n_T \frac{1}{\tau'_T} \xrightarrow{equilib} 0$$

**(eqns. 2)**

describes the photo excitation where no direct excitation of the triplet state occurs, and $I_{photo}$ is the excitation by light absorption.

In the following, we will make the coarse assumption that the internal rate constants of the model (Supporting Figure 1) are the same for phosphorescence and fluorescence and that the non-radiative decay of the $S_1$ state is negligible because the radiative decay dominates due to the Purcell effect in the STM.

Experimental observation of the steady state cannot access the time constants in the model directly, but allow to observe the triplet to singlet intensity ratios in electroluminescence $R_{el} = \frac{\eta_T}{\eta_S}\frac{n_{elT}}{n_{elS}}\frac{\tau_{rS}}{\tau_{rT}}$ and in photoluminescence $R_{pl} = \frac{\eta_T}{\eta_S}\frac{n_{plT}}{n_{plS}}\frac{\tau_{rS}}{\tau_{rT}}$. Here $\eta_S$ and $\eta_T$ are the detection efficiencies for singlet and triplet emission.

Using these definitions together with eqns. 1, eqns. 2 and the above assumptions, we find that experimental detection efficiencies, the excitation strengths ($I_{el}$ and $I_{photon}$) and the triplet state lifetime cancel out and we have:

$$\Delta := \frac{R_{el}}{R_{pl}} = \left(3\ \frac{\tau_{ISC}}{\tau_{rS}}\ + 4\right) \qquad (eq. 3)$$

From the experimentally observed triplet to singlet ratios discussed in the main text, we have $\Delta \approx 34$ and can derive:

$$\frac{\tau_{ISC}}{\tau_S} \approx \frac{\Delta - 4}{3} \approx 10 \qquad (eq. 4)$$

and

$$\frac{n_T}{n_S} \approx 10\ \frac{\tau_S}{\tau_T} \qquad \text{(eq.5)}$$

Using the experimentally observed line width (2.49 meV) we obtain a coarse estimate of the singlet state life times of $\tau_S \approx 0.3$ ps which is of the same order as the literature value $\tau_S \approx 0.7$ ps derived by dynamic photon-correlation studies of Zn-Phthalocyanine.[20] We then find for the intersystem crossing $\tau_{ISC} \approx 3$ ps comparable to the experimental value for Pd-Phthalocyanine of $\tau_{ISC} \approx 1.5$ ps.[21]

## SUPPORTING FIGURES

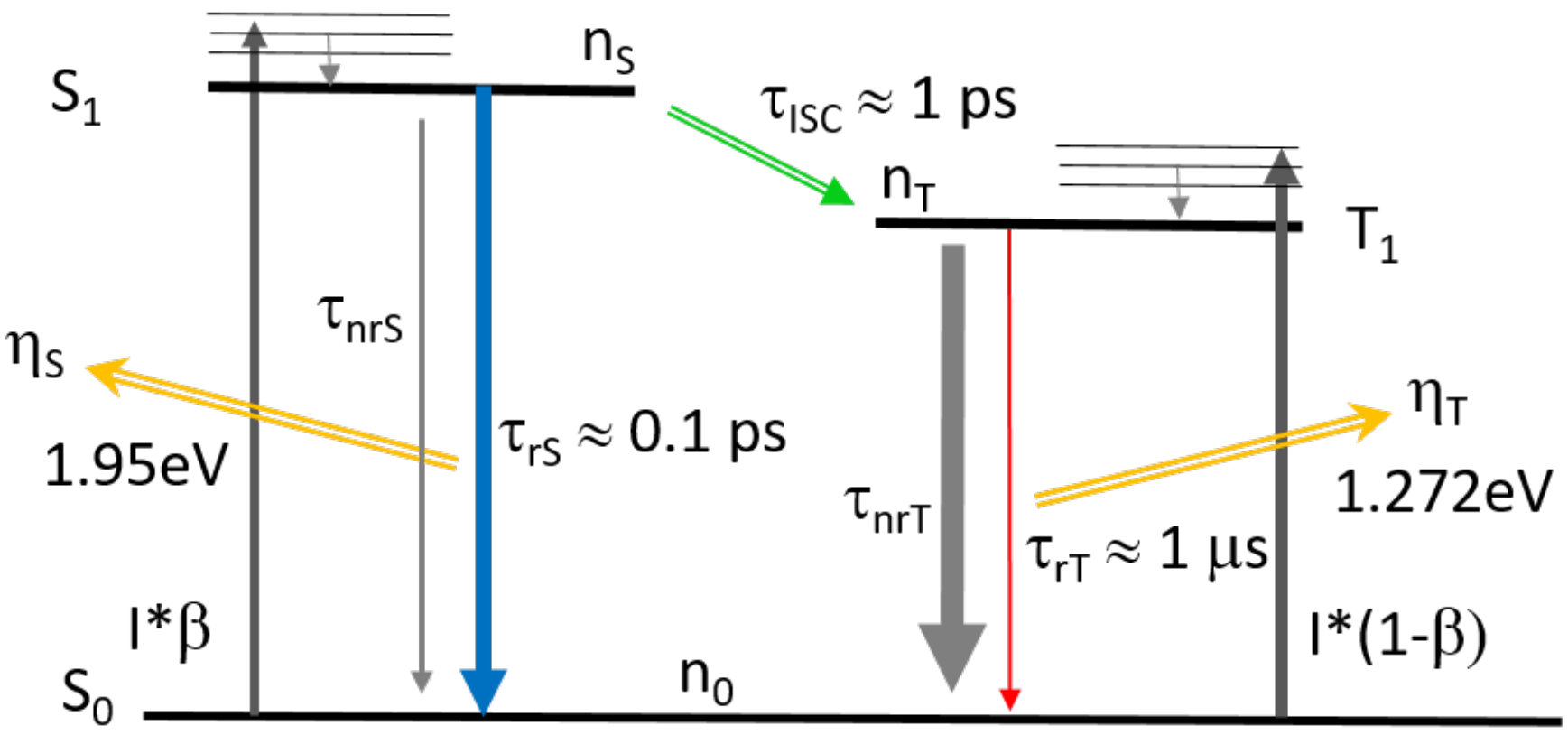

**Supporting Figure 1. Model for fluorescence and phosphorescence** introducing the variables used for the calculation. $n_0$ denotes the occupation of the $S_0$ ground state of the system. $n_S$ and $n_T$ denote the occupation of the lowest singlet ($S_1$) and lowest triplet ($T_1$) state, respectively. The life times, that are the inverse rate constants, are denoted by $\tau$ with the index denoting *nr* for non-resonant, *r* for resonant, and ISC for inter-system-crossing transitions. $\eta_S$ and $\eta_T$ are efficiency factors comprising the coupling of the respective transition to the far field, the intrinsic instrumental transmission and the detector efficiency. $\beta$ controls the branching ratio between excitation of the singlet ($\beta$) and excitation of the triplet (1-$\beta$). The overall excitation efficiency of the system is also assumed to be included as a factor in the detection efficiencies $\eta$.

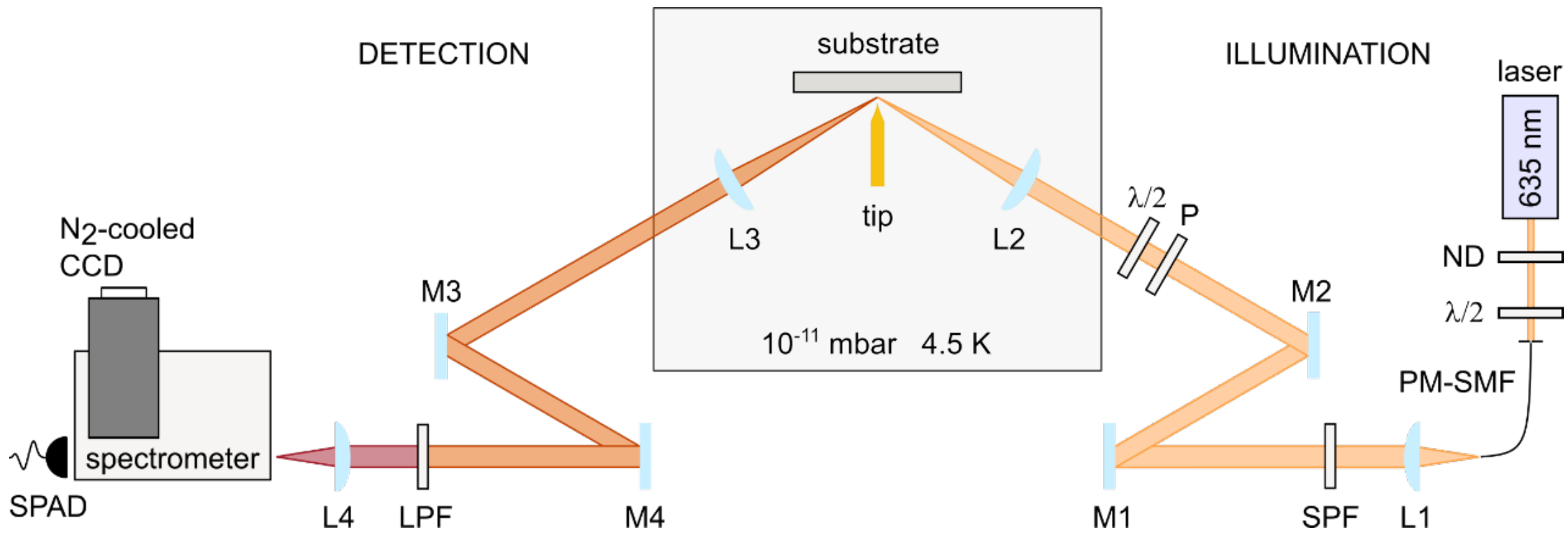

**Supporting Figure 2. Experimental setup.** Schematic illustration of the ultrahigh vacuum (UHV) low-temperature scanning tunneling microscope (Omicron) operating at 4.3 K. All experiments have been performed at the Surface and Interface Science Laboratory, RIKEN, Japan. The STM stage is equipped with two optical lenses (each covering a solid angle of $\sim 0.5$ sr). For the STML measurement, the emitted light was collimated by lens L3 and directed out of the UHV chamber, where it was refocused onto a grating (50, 300, or 1200 grooves $mm^{-1}$) spectrometer (Acton, SpectraPro 2300i) and detected with a charge coupled device (CCD) (Princeton, PyLoN:100) cooled with liquid nitrogen. A path switching mirror in the spectrometer allows to direct the light to an avalanche photo diode (APD: Excelitas SPCM-AQRH) which is used to obtain the current dependence of light emission. The pulses from the APD were counted using a multi-channel DAQ device (National Instruments) and monitored using a LabVIEW software package. For the TEPL measurement, excitation was induced using a laser diode (Thorlabs). Neutral-density filters (ND) were employed to control the laser power. The laser light was coupled to a polarization-maintaining single-mode fiber (PM-SMF: Thorlabs) and collimated by lens L1. Then the laser beam passes through a short pass filter (SPF: Semrock) to clean up the spectral region to be measured. Polarization is defined by a polarizer (p) and a half wave plate (λ/2). In this study, *p*-polarization is used. Finally, the laser is focused into the STM junction by lens L2. The emitted light is collimated by lens L3 and directed out of the UHV chamber, where it passes through a long pass filter (LPF: Semrock) to block the excitation laser light (the LPF is removed during STML measurements). The laser energy is tuned by controlling the temperature of the diode and monitored using a grating (1800 grooves $mm^{-1}$) spectrometer (Acton IsoPlane-320) with a nitrogen-cooled CCD photon detector (Princeton, Spec10).

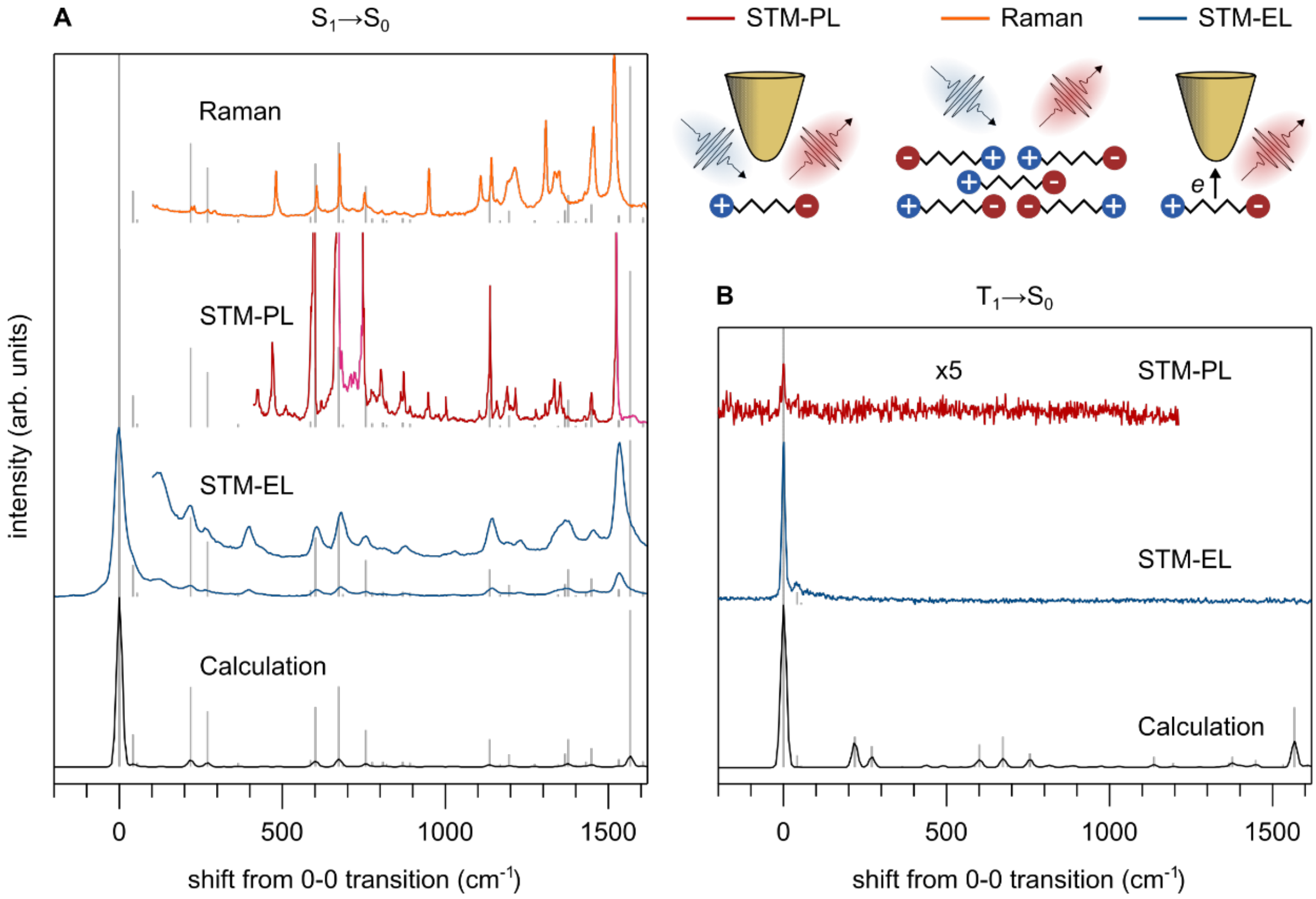

**Supporting Figure 3. Comparison of vibronic satellites for fluorescence and phosphorescence.** (A) Top to bottom: Raman spectrum for PtPc powder using a 532 nm laser (grating: 600 grooves mm$^{-1}$), TEPL spectrum next to a PtPc molecule atop 4 ML NaCl on Ag(111) ($I$ = 3 pA, $V$ = 1 V; laser power: 1 μW, $t$ = 30 s, grating: 1200 grooves mm$^{-1}$), STML spectrum for the PtPc molecule obtained with the tip placed atop a HOMO lobe ($I$ = 60 pA, $V$ = −2.6 V; $t$ = 120 s, grating: 300 grooves mm$^{-1}$), and calculated vibronic spectrum of a neutral PtPc molecule in the gas phase for the $S_1 \rightarrow S_0$ transition. (B) Top to bottom: TEPL ($t$ = 120 s), STML, and calculated optical spectrum of a neutral PtPc molecule for the $T_1 \rightarrow S_0$ transition.

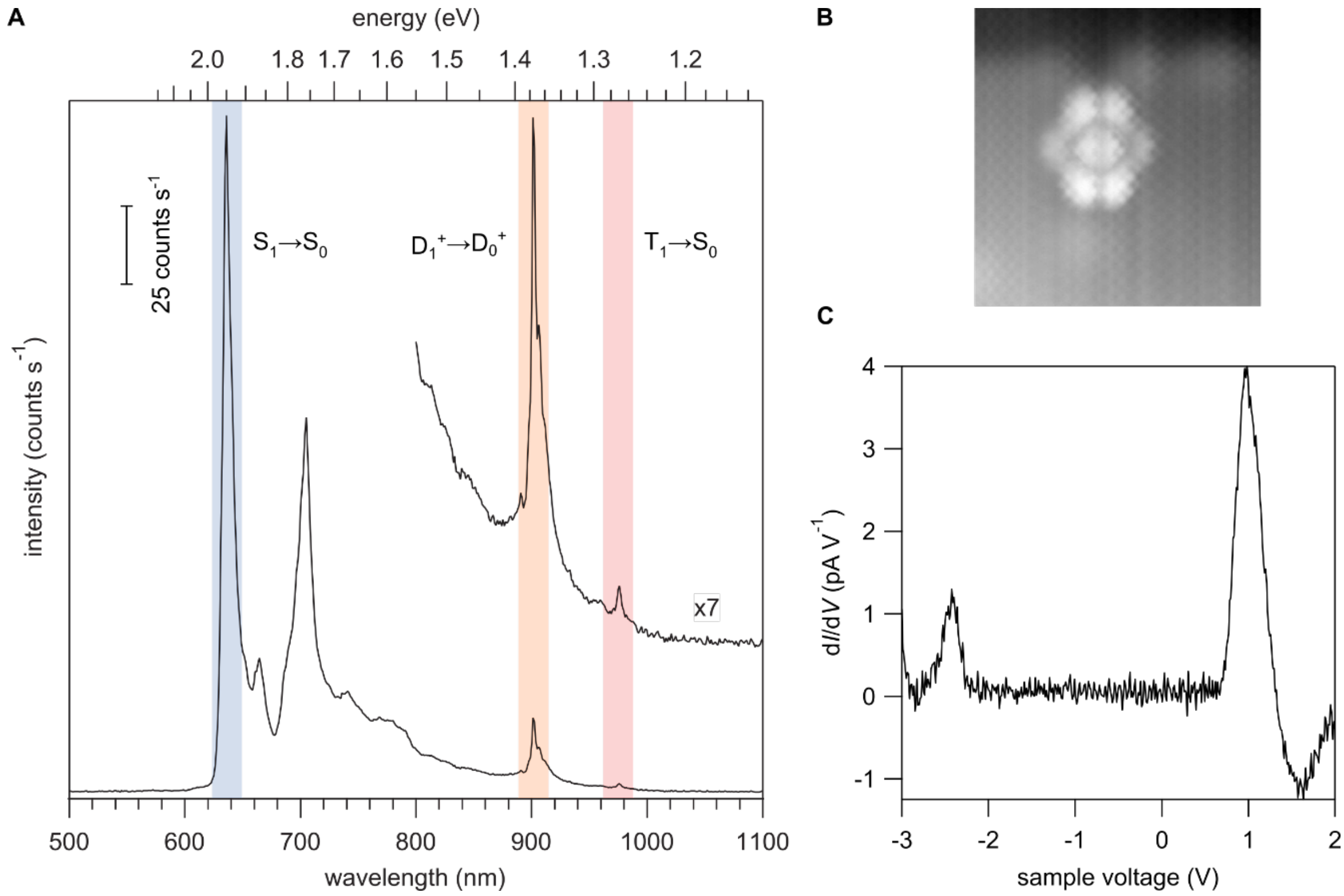

**Supporting Figure 4. Emission from charged PtPc molecule.** (A) STML spectrum obtained for a charged PtPc species ($PtPc^+$) adsorbed atop 3 ML NaCl on Ag(111) at negative sample voltage ($I$ = 200 pA, $V$ = -2.7 V, $t$ = 200 s, grating: 50 grooves $mm^{-1}$). These STML spectra have not been normalized by the plasmon spectra in contrast to the other spectra of the study. (B) STM topography image of the molecule showing degeneracy lifting of the LUMO and LUMO+1 orbital ($I$ = 4 pA, $V$ = 1.1 V, size: 5×5 $nm^2$). (C) d$I$/d$V$ spectrum obtained atop the center of the molecule showing the molecular frontier orbitals, similar to the case where only neutral emission is observed (see Fig. 1A).

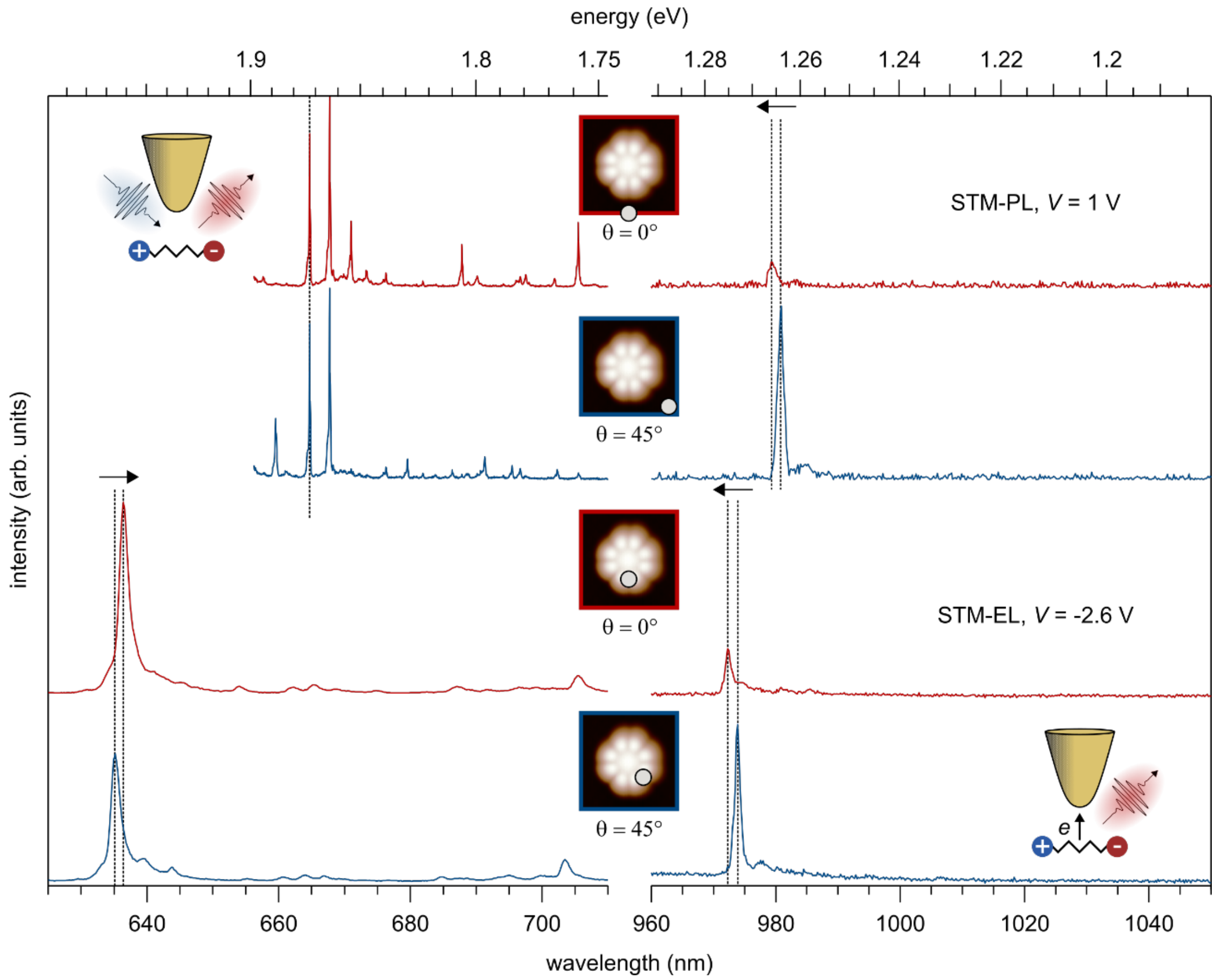

**Supporting Figure 5. Angle dependence of fluorescence and phosphorescence.** Bottom – STML spectra showing $S_1$ (left) and $T_1$ emission (right) obtained for azimuths $\theta$ = 0° and $\theta$ = 45°. Distance from the molecule $r$ = 1.1 nm ($I$ = 80 pA, $V$ = −2.6 V; for $S_1$: $t$ = 30 s, and for $T_1$: $t$ = 120 s, grating: 300 grooves mm$^{-1}$). Top – TEPL spectra showing vibronic peaks (left) and $T_1$ emission (right) for $r$ = 1.9 nm, i.e., off the molecular orbital ($I$ = 3 pA, $V$ = 1 V, $t$ = 300 s, grating: 300 grooves mm$^{-1}$).

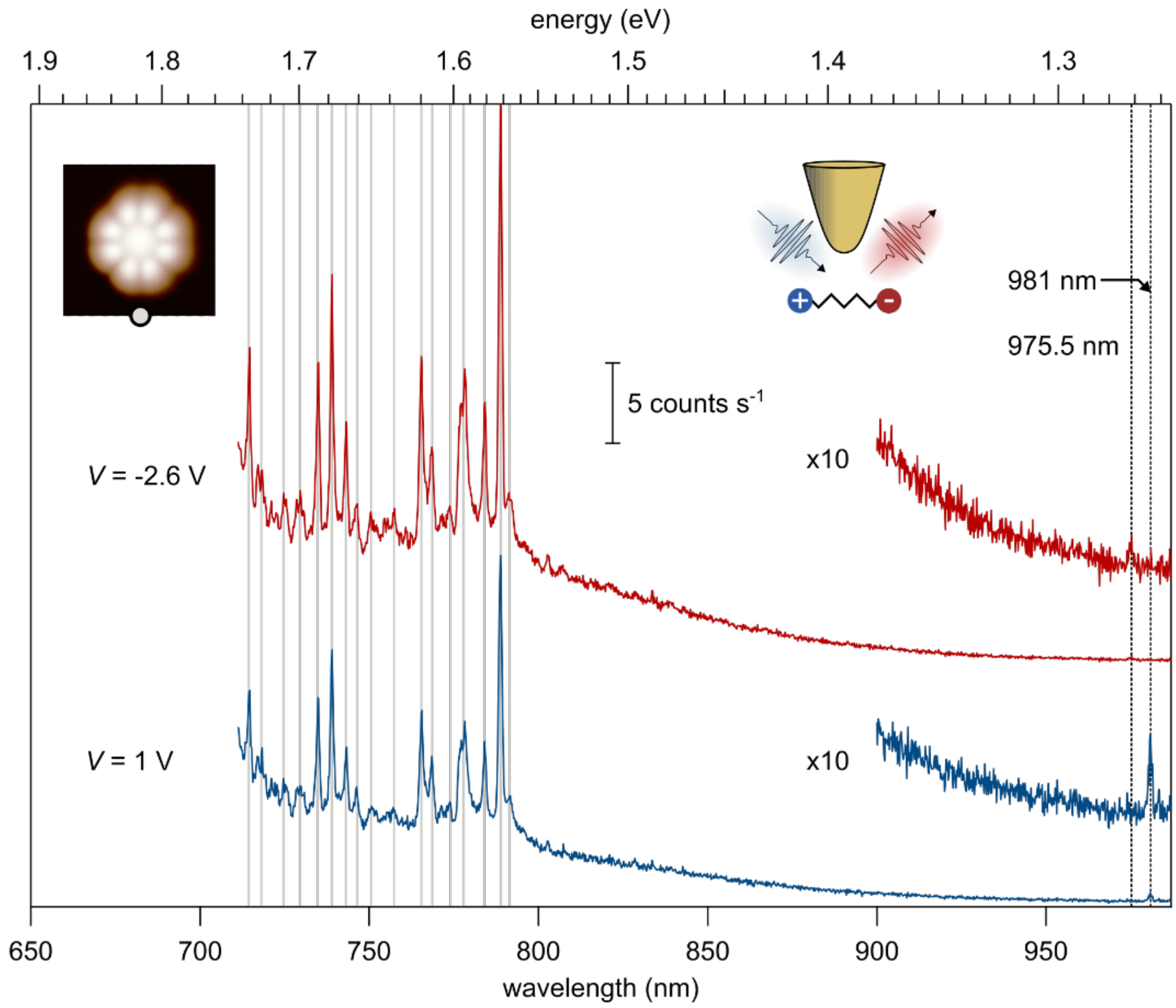

**Supporting Figure 6. Extended TEPL spectra at different voltage polarities demonstrating the Stark shift of $T_1$ emission.** Top – TEPL spectra at sample voltage $V$ = −2.6 V (top) and +1 V (bottom) showing vibronic peaks and the $T_1$ emission line (laser power: 1μW, $I$ = 3 pA, $t$ = 5 min, grating: 300 grooves mm$^{-1}$). The blowups (×10) of the $T_1$ emission line demonstrate a voltage dependent redshift. Inset (top left): STM topography image of the molecule with a gray dot marking the tip-position ($I$ = 3pA, $V$ = −2.6V, size: 3.5×3.5 nm$^2$).

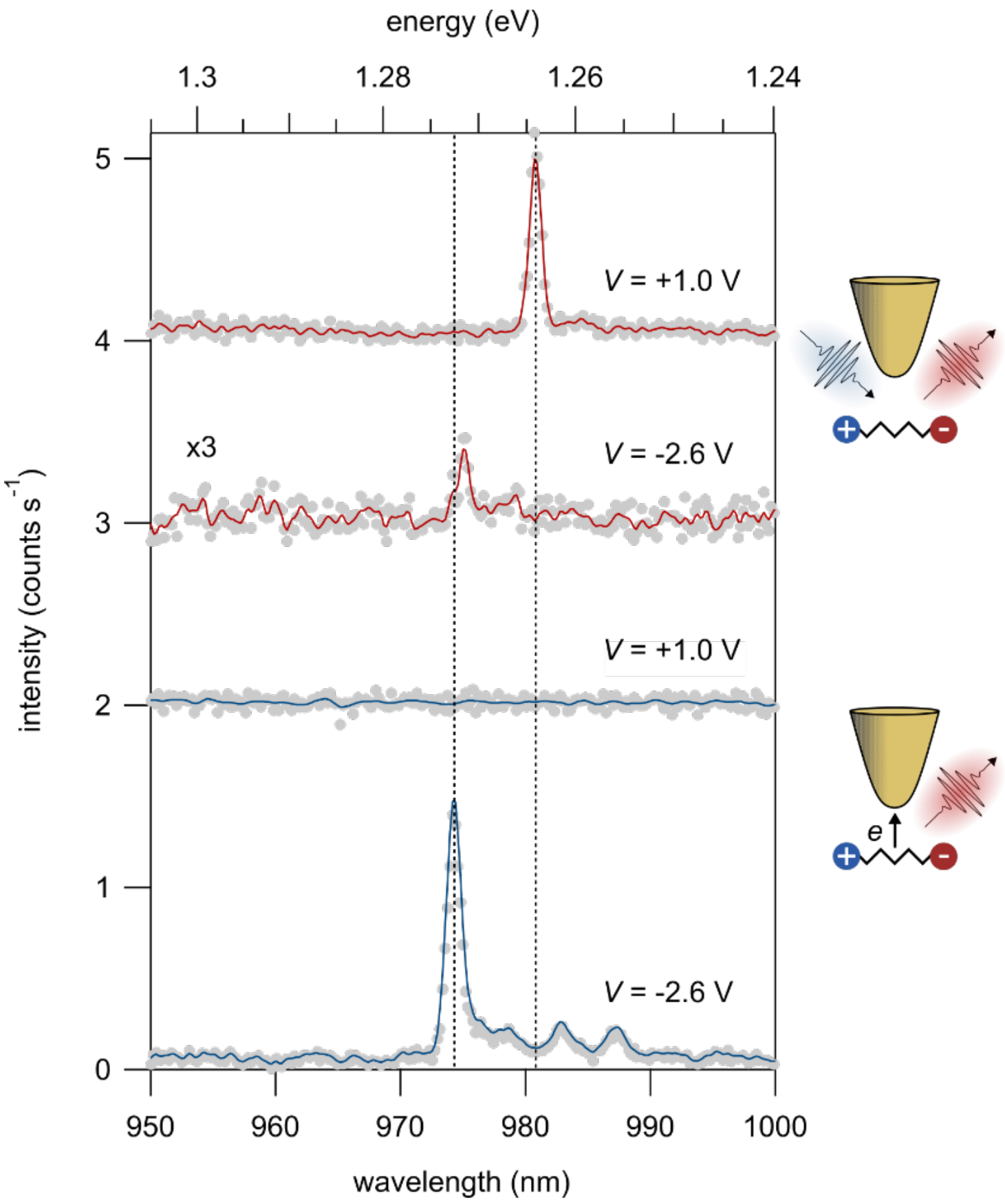

**Supporting Figure 7. Voltage dependence of the $T_1$ emission line comparing STML and TEPL.** Bottom – STML spectra at $V = -2.6$ V and +1 V ($I$ = 60 pA, $t$ = 120 s, $\theta$ = 0°, $r$ = 1.1 nm, grating: 300 grooves mm$^{-1}$). Top – TEPL spectra at $V = -2.6$ V and +1 V (laser power: 1 μW, $I$ = 3 pA, $t$ = 300 s, $\theta$ = 0°, $r$ = 1.9 nm, grating: 300 grooves mm$^{-1}$) showing the voltage dependent shift of the $T_1$ line.

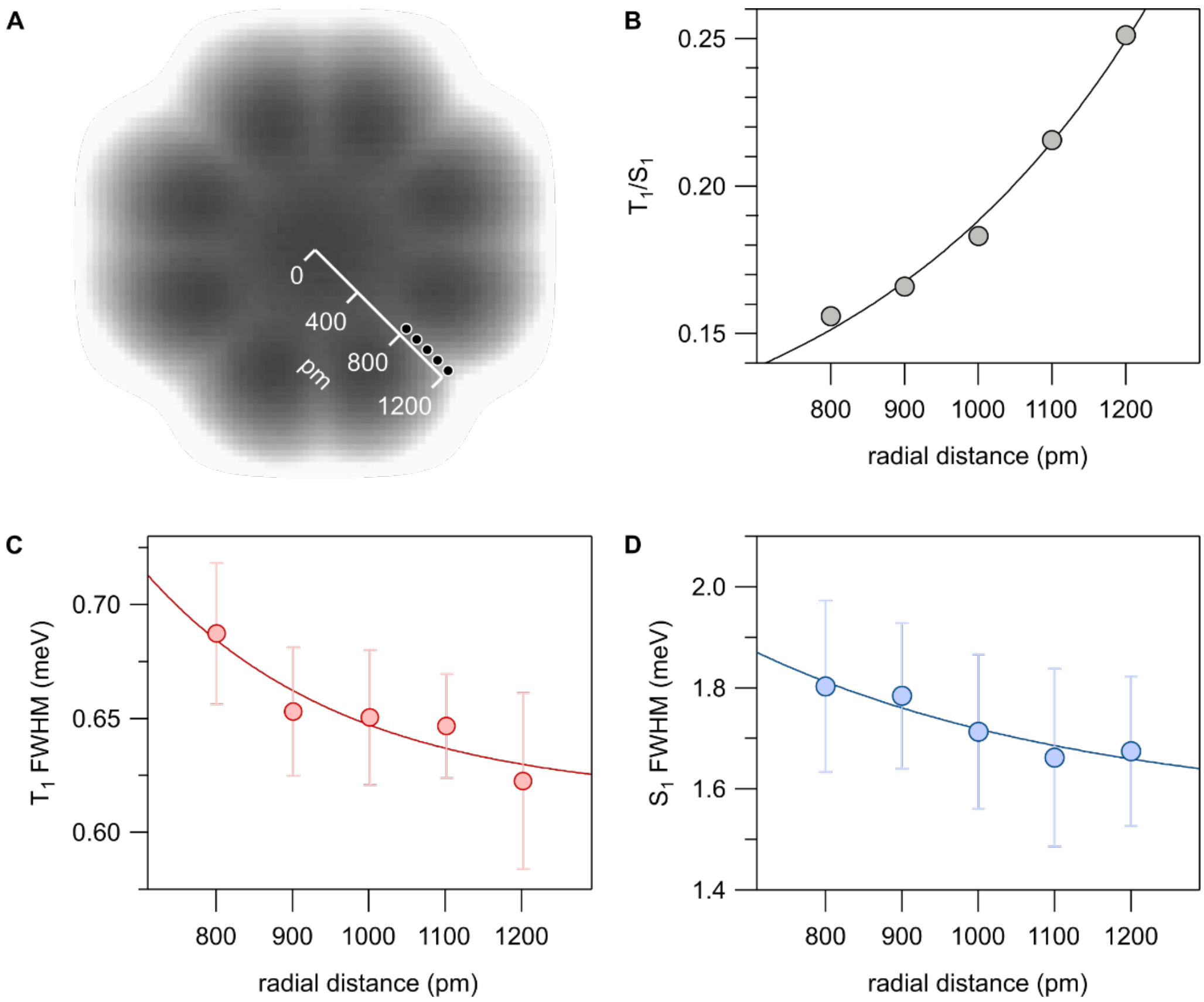

**Supporting Figure 8. Radial distance-dependent X-P coupling (Purcell effect) of the $S_1$ and $T_1$ emissions.** (A) STM topography of the molecule with tip positions marked for the data presented in B-D ($I$ = 4 pA, $V$ = -2.6 V). (B) $T_1/S_1$ intensity ratio as a function of radial distance from the molecule center ($I$ = 50 pA, $V$ = -2.6 V; for $S_1$: $t$ = 10 s, and for $T_1$: $t$ = 120 s; grating: 300 grooves $mm^{-1}$). (C, D) FWHM evaluated using a Lorentzian line shape fit to the $T_1$ and $S_1$ emission lines, respectively, for the same tip positions as in B. All solid lines are exponential fits to guide the eye.

**REFRENCES**